\documentclass[12pt]{article}
\usepackage{graphicx}
\usepackage{caption}
\usepackage{subcaption}
\usepackage[utf8]{inputenc}
\usepackage[toc,page]{appendix}
\usepackage{amsmath}
\usepackage{tensor}
\usepackage{gensymb}
\usepackage{multicol}
\usepackage{a4wide,graphics,amsmath,amssymb,cite,nicefrac}
\usepackage[verbose]{wrapfig}
\usepackage{authblk,hyperref}
\usepackage{url,amsfonts}
\usepackage{physics}
\usepackage{xcolor}
\numberwithin{equation}{section}
\begin{document}
	\begin{titlepage}

		\vskip 1.0cm

	\begin{center}
			{\boldmath \Huge{Self-Supporting Wormholes in Four Dimensions with Scalar Field}}

		\vskip 2.0cm
		
	{\bf \large Ankit Anand }
		\footnote{ E-mail : Anand@physics.iitm.ac.in}
		\vskip 0.5cm
		{\it
		
		Department of Physics, Indian Institute of Technology Madras, \\
Chennai 600 036, India
			}\\
		
		\vskip 10pt
			\end{center}
	\begin{center}
		    
        \end{center}
		
		\vskip 1.5cm
		
		\begin{center}
			
			 {\bf ABSTRACT}
			
			 \end{center}%
			 
	In this paper, we investigated the space-time obtained by quotients of the $AdS_4$ space-time. Further quotient with specific $\mathbb{Z}_2$ is considered. Taking up to the first-order perturbation in metric, we estimated the backreaction of the matter field on space-time geometry. We can calculate the expectation value of stress-energy tensor  by pulling it back onto the covering space. The average null energy becomes negative when the suitable boundary condition is chosen, resulting in a traversable wormhole.

	\end{titlepage}

%\maketitle

\section{Introduction}
Wormholes have been a topic of interest for both scientists \cite{Einstein:1935tc} \cite{Morris:1988tu} and the general public as it provides a way for rapid transit between two distant points in space or also for communication over long distances. Wormholes are Einstein equation solutions that use a throat to connect two otherwise different space-times or two widely separated areas of the same space-time. Classically, wormholes are not traversable, meaning that a causal curve cannot pass through the wormhole's throat, connecting the two different regions. A traversable wormhole is possible only if the geodesics entering the wormhole on one side (and thus converging as they approach the throat) will emerge on the other side, diverging from each other. Raychaudhuri’s equation showed that it can only happen if certain energy conditions are violated - the null energy condition (NEC) and the averaged null energy conditions (ANEC). The ANEC asserts that, there must be an infinite number of null geodesics with a tangent vector $k^\mu$ and affine parameter $\lambda$ passing through the wormhole to satisfy the condition
\begin{equation}
    \int_{-\infty}^\infty T_{\alpha \beta} k^\alpha k^\beta d \lambda < 0 \ .
\end{equation}
 The ER=EPR conjecture \cite{Maldacena:2013xja}  states that whenever two particles are entangled, they must be connected through a wormhole. \cite{Morris:1988cz} originally addressed the issue of wormhole traversability for static, spherically symmetric wormholes, and it has been shown that wormholes must have exotic matter for traversability. \cite{Hochberg:1998ha,Morris:1988tu, Visser:2003yf} further investigated this issue and established the violation of the average null energy condition (ANEC) as an essential requirement for wormhole traversability. The ANEC has been demonstrated to hold for achronal null geodesics \cite{Graham:2007va,Kelly:2014mra,Wall:2009wi}. As a result, space-times with only achronal null geodesics do not allow for traversable wormholes. The topological censor theorem \cite{Friedman:1993ty} and its generalization to asymptotically localised anti-de Sitter spaces \cite{Galloway:1999bp} declare that any causal curve whose end points reside in the boundary at infinity $(\mathcal{I})$ can be transformed to a causal curve that wholly lies in $(\mathcal{I})$.
\par
Gao, Jafferis, and Wall \cite{Gao} recently made a significant advance in this direction. By adding a time-dependent coupling between the two asymptotic regions of an eternal BTZ black hole, they were able to create a traversable wormhole. Using the point splitting method, they calculated the one loop stress energy tensor. By correctly choosing the sign of the coupling, the vacuum expectation value of the double null component of the stress energy tensor may be made negative, allowing the wormhole to be traversable. These findings were then generalised in \cite{Caceres:2018ehr} to investigate the effect of rotation on wormhole size. In \cite{Maldacena:2018lmt}, a connection between the two boundaries was used to create an eternally traversable wormhole in nearly-$AdS_2$ space-time. 
\par
%%%%%%%%
In the presence of massless fermions, \cite{Maldacena:2018gjk} constructed a four-dimensional traversable wormhole by connecting the throats of two charged extremal black hole geometries. This construction did not rely on any non-local external coupling between the two boundaries, and the result was what are known as self-supporting wormholes, which form purely from the local dynamics of the fermion fields existing in the bulk of space-time. In \cite{Bintanja:2021xfs} authors have constructed eternal AdS$_4$ traversable wormhole by coupling CFT$_3$ boundary theories. 
 In  \cite{Scalar} authors have constructed the traversable wormhole without adding any coupling between its asymptotic regions. They have presented an alternative analysis to ascertain traversable wormholes from bulk dynamics by considering a free scalar field in quotients of $AdS_3$ and $AdS_3\cross S^1$ by discrete symmetries. The authors calculated the gravitational back reaction and demonstrated that causal curves that cannot be deformed to the boundary exist in space-time. Taking the quotient by a discrete symmetry is essential in that it destroys the globally specified Killing field, which is crucial for attaining the average null energy condition. This finding was later extended \cite{fermions} to include fermions in the bulk and \cite{spin 1} to include massive spin one, resulting in traversable wormholes.
\par
In this paper, we generalise the aforementioned conclusions to the four dimensions. In the case of $AdS_4$, an exact analytic equation for the propagator in closed form is not possible. We computed the expectation value of the stress tensor by fixing different values of mass $m$ and show that this leads to traversable wormholes when sufficient boundary conditions are imposed. The preliminaries for developing self-supporting wormholes from free scalar fields are summarised in the following section. By quotienting out the AdS4, we get the space-time in \ref{3+1}. The expectation value of the stress tensor for the scalar field is then computed using the images approach. The choice of co-ordinate is discussed in the appendix \ref{Kruskal} and the linearized Einstein's equation upto first order is discussed in appendix \ref{Linearised}.

%%%%%%%%%%%%%%%%%%%%%%%%%%%%%%%%%%%%%%%%%%%%%%%%%%%%%%%%%%%%%%%%%%%%%%%%%%%%%%%%%%%%%%%%%%%%%%%%%%%%%%%%%%%%%%%%%%%%%%%%%%%%%%%%%%% 
\section{Preliminaries}\label{2+1}
The AdS$_3$ metric in Kruskal-like coordinates $(U,V,\phi)$ is 
\begin{equation}\label{gUV}
    dS^2 = g_{\alpha \beta}dx^\alpha dx^\beta = \frac{1}{(1+UV)^2}\Big(-4 l^2 dU dV + r_+(1-UV)^2 d\phi^2\Big)\ ,
\end{equation}
where $\phi$ is azimuthal angle. By identifying $\phi \thicksim \phi+2\pi$, it give rise to a non-rotating BTZ black hole with horizon radius $r_+$. $U=0$ and $V=0$ are the horizons of black hole and $1+UV=0$ indicates the boundary of black hole. The RP2 -geon \cite{Louko:1998hc} is generated by multiplying this geometry by the $\mathbb{Z}_2$ isometry $J$, which has the following effect on the co-ordinates: $J: (V, U, \phi ) \rightarrow (V, U, \phi+\pi )$. The above solution was constructed and discussed in \cite{Banados:1992wn}. From the gauge-gravity perspective, the solution has been discussed in \cite{Maldacena:2001kr}. In the case of wormhole discussed in \cite{Aminneborg:1997pz}. This is the first solution in which the authors of \cite{Gao}  have discussed the traversability issue of the wormhole by taking a appropriate double trace deformation coupled to two boundaries. The appropriate double trace deformation in boundary CFT amounts to adding a stress tensor in bulk, resulting in a perturbation of the space-time geometry 

\par 
We'll start with the fact that is discussed in \cite{Gao}. By using the fact that the background metric \eqref{gUV} has constant $g_{UV}$ along the horizon $(V = 0)$, which implies that the geodesic equation at linear order implies a null ray starting from the right boundary in the far past to have 
\begin{eqnarray}\label{geodesics}
    V(U) &=& -(2g_{UV}(V=0))^{-1} \int_{-\infty}^U dU h_{UU} \ , \\
         &=& \frac{1}{2 l^2} \int_{-\infty}^U dU h_{UU} \ ,
\end{eqnarray}
where $h_{kk}$ is the norm of $k^a$ after first-order back-reaction from the quantum stress tensor.  By taking one of the horizon into account, let's take $V=0$ horizon with the horizon generator $k^\lambda$ such that $k^\lambda\partial_\lambda = \partial_U$. The null geodesics tangent to this horizon can be parametrized by choosing $U$ as the affine parameter. For the metric perturbation on the chosen horizon ($V=0$), the linearized Einstein's equation for $h_{\mu \nu}=\delta g_{\mu \nu} \thicksim \order{\epsilon}$ is written as \footnote{Detailed discussion in appendix \ref{Linearised}.}
\begin{equation}\label{Linearized}
    \frac{1}{2} \Big[\frac{1}{l^2}(h_{UU}+\partial_U(Uh_{UU}))-\frac{1}{r_+^2}\partial_U^2h_{\phi\phi}\Big]= 8 \pi G_N T_{UU} \ ,
\end{equation}
By integrating the equation \eqref{Linearized} over all $U$ to get the shifts in the ray from far past to far future. While integrating and using the asymptotically AdS boundary conditions the equation reduces to 
\begin{equation}
    8\pi G_N \int dU \expval{T_{kk}} = \frac{1}{2l^2} \int dU h_{kk} \ ,
    \end{equation}
    the shift at far future is 
    \begin{equation}
          \Delta V(+\infty) = -\frac{8 \pi G_N l^2}{g_{UV}(0)} \int_{-\infty}^\infty dU \expval{T_{kk}} = 4 \pi G_N \int_{-\infty}^\infty dU \expval{T_{kk}} \ .
    \end{equation}
    The time delay of the null geodesics starting from $U = -\infty$ and ending at $U=\infty$ can be measured using the quantity $\Delta V(\infty)$. This quantity also provides a measure for the size of the wormhole's opening. The wormhole becomes traversable iff  the ANEC is violated or, equivalently $\Delta V(\infty) < 0$. By choosing an appropriate non-local coupling between the boundaries, it has been shown in \cite{Gao} that taking a one-loop stress tensor can violate the ANEC and results in wormhole traversability.
    \par
    The above construction relies on the addition of any non-local boundary interaction. Another method has been proposed in \cite{Scalar} to give rise to a traversable wormhole without adding non-local coupling. This method relies on choosing a suitable $\mathbb{Z}_2$ quotient of BTZ black hole space-time$\tilde{M}$; it results in a smooth, globally hyperbolic manifold, called as $\mathbb{RP}^2$-geon \cite{Louko:1998hc}, $M$. The manifold $\tilde{M}$ is also called the covering space. A new homotopy cycle in manifold $M$ arises due to the introduction of the $\mathbb{Z}_2$ quotient and, it allows to take the scalar field in $M$ to be either periodic or anti-periodic around this circle. By using the method of images, one can relate the state on $M$ and $\tilde{M}$. The points $\tilde{x} \in \tilde{M}$ can be projected into $M$ by taking an isometry, let's say $J$, i.e., the pairs $(\tilde{x}, J\tilde{x})$ project on point $x \in M$. By using the Method of images, the scalar quantum fields $\tilde{\phi}(x)$ in $\tilde{M}$ are used to construct the quantum fields in $M$ as
    \begin{equation} \label{PAP condition}
        \phi_\pm(x) = \frac{1}{\sqrt{2}} \left(\tilde{\phi}(\tilde{x}) \pm \tilde{\phi}(J\tilde{x})\right) \ ,
    \end{equation}
    where $\pm$ corresponds to the periodic and anti-periodic boundary condition. The points $\tilde{x}$ and $J\tilde{x}$ can't coincide as $M$ is smooth. Thus they are spacelike separated and the quantum fields at these points commute.
    \par
    The action for free scalar field $\phi_\pm(x)$ in M is
    \begin{equation}
	    S=\int d^4x \sqrt{-g}\left(-\frac{1}{2}g^{\alpha \beta}\partial_\alpha \phi_\pm(x) \partial_\beta \phi_\pm(x) -\frac{1}{2}m^2 \phi_\pm^2(x) \right) \ .
	\end{equation}
	The stress energy tensor by varying the action with respect to $g^{\alpha \beta}$, 
	\begin{equation} \label{stress energy}
    T_{\alpha \beta} = \partial_\alpha \phi \partial_\beta \phi -\left[\frac{1}{2}g_{\alpha \beta} g^{\gamma \delta} \partial_\gamma \phi \partial_\delta \phi +\frac{1}{2} g_{\alpha \beta}m^2 \phi^2\right] \ .
\end{equation}
To compute the expectation value of double null component of stress tensor in Hartle-Hawking state. Hartle-Hawking state in $M$ is represented as $\ket{HH,M}$ and in $\tilde{M}$ as $\ket{HH,\tilde{M}}$. As the quantity of interest is $k^\alpha k^\beta T_{\alpha \beta}$, the term inside the parenthesis in equation \eqref{stress energy} vanishes because of $T_{\alpha \beta}k^\alpha k^\beta =0$. Finally we have \cite{spin 1}
\begin{eqnarray} \label{expac}
\bra{HH,M}k^\alpha k^\beta T_{\alpha \beta \pm} \ket{HH,M} &=& \pm \bra{HH,\tilde{M}}k^\alpha \partial_\alpha \tilde{\phi}(\tilde{x}) k^\beta \partial_\beta \tilde{\phi}(J\tilde{x}) \ket{HH,\tilde{M}}  \nonumber \\
                                &=& \pm \bra{HH,\tilde{M}}\partial_U \tilde{\phi}(\tilde{x}) \partial_U \tilde{\phi}(J\tilde{x}) \ket{HH,\tilde{M}} \ .
\end{eqnarray}
This result emphasizes the main idea. Unless the integral of the right-hand side disappears, it will be negative for some boundary conditions $(\pm)$. Backreaction will then make the wormhole traversable with that decision. It is thus only necessary to investigate this integral in certain circumstances, demonstrating that it is non-zero and quantifying the degree to which the wormhole becomes traversable. This has been calculated for several smooth, globally hyperbolic, $\mathbb{Z}_2$  quotients of $ \rm BTZ$ and $BT Z \times S^1$ space-times \cite{Scalar,fermions,spin 1}. The results have been subsequently utilized to demonstrate the ANEC violation.

%%%%%%%%%%%%%%%%%%%%%%%%%%%%%%%%%%%%%%%%%%%%%%%%%%%%%%%%%%%%%%%%%%%%%%%%%%%%%%%%%%%%%%%%%%%%%%%%%%%%%%%%%%%%%%%%%%%%%%%%%%%%%%

\section{BTZ black hole in 3+1 dimensions} \label{3+1}
A BTZ black hole is a space-time obtained by identifying points in AdS-space. The BTZ black hole could be reviewed as the quotient space $[AdS]/\mathrm{G}_T$. $\mathrm{G}_T$ is a group generated by $\Gamma$ : $\mathrm{G}_T=\{\Gamma^n; n \epsilon \mathbb{Z}\}$, $\Gamma=e^{\alpha \xi}$ for some fixed $\alpha$, $\Gamma$ represents the discrete symmetry of AdS space and $\xi$ is the killing field. If $\xi$ is timelike in some regions of AdS-space then point identified by $e^{\alpha \xi}$ results in closed timelike curve(CTC). So, an observer avoid to entering in the region where $\xi$ is timelike. When $\xi$ is lightlike i.e., $\xi_\mu \xi^\mu=0$ hypersurface  is known as \textit{singularity} and interpreted as horizon. 
\par 
Let's start with defining $3+1$ AdS space as hyperboloid
\begin{equation} \label{Hyperboloid}
    -T_1^2-T_2^2+X_1^2+X_2^2+X_3^2 = -l^2 \ ,
\end{equation}
embedded in the flat $5-$dimensional space with metric
\begin{equation}
    ds^2=-dT_1^2-dT_2^2+dX_1^2+dX_2^2+dX_3^2 \ .
\end{equation}
This surface given by the above equation, in particular, has a Killing vector $\xi^\alpha \partial_\alpha= \frac{r_+}{l}(T_1 \partial_{X_1}+ X_1 \partial_{T_1})$, which is a boost in the $(T_1,  X_1)$ plane with a norm of $\xi^2=\frac{r_+^2}{l^2}(T_1^2-X_1^2)$, where $r_+$ is an arbitary real constants. The Norm can be positive, negative or zero. That defines the existance of Black hole. Locating points along the orbits of $\xi^\alpha$ is necessary to build a black hole. The orbits of are timelike in the region $\xi^2 < 0$. They will, however, closed (i.e., they contains closed timelike co-ordinates) after the identification has been made. As a result, the region $\xi^2 < 0$ is not physical in this sense and no longer a physical and its boundary $\xi^2 = 0$ is singular. As a result, there are three regions in spacetime that are of the interest : $I := r_+ < \xi^2 < \infty$, $II :=0 < \xi^2 < r_+$ and $III := -\infty < \xi^2 \leq 0$. The causal structure has been discussed in \cite{Banados:1998dc} \cite{Holst:1997tm} \cite{Aminneborg:2008sa} . 
 \par 
 From \cite{Guica:2014dfa}, for non-rotating BTZ black hole is obtained by restricting to the region $T_1^2>X_1^2$ (i.e. ,$\xi^2>0)$, where the Killing vector is space-like, and quotienting by the discrete isometry group results in black hole solution. By introducing local coordinates of AdS space in the region $\xi^2 > 0$ to write down the identification along the orbits of $\xi^\alpha$ explicitly as
\begin{eqnarray} \label{coordinate choice}
 T_1 &=& \frac{l r}{r_+} \text{Cosh} \left(\frac{r_+ \phi}{l}\right) \nonumber \\
 X_1 &=& \frac{l r}{r_+} \text{Sinh} \left(\frac{r_+ \phi}{l}\right) \nonumber \\
 X_i &=& \frac{2l y_i}{1-y^2} \ ,
\end{eqnarray}
with 
\begin{equation} \label{r r+ y}
    r=r_+\frac{1+y^2}{1-y^2} , \hspace{1cm} y^2=-y_0^2+y_2^2+y_3^2
\end{equation}
where $X_i's$ are $T_2$, $X_2$ and $X_3$ for $y_0$, $y_2$ and $y_3$ respectively. Here $-\infty < y_i <\infty$ and $-\infty < \phi <\infty$ with restriction $-1 < y^2 <1$. The boundaries i.e., $r \rightarrow \infty$ represents a hyperbolic ''ball" $y^2=1$. Induced metric can be written as
 \begin{equation}
     ds^2 = \frac{l^2(r^2+r_+^2)^2}{r_+^2} \left( -dy_0^2+dy_2^2 +dy_3^2 \right) + r^2 d\phi^2 \ .
 \end{equation}
 The killing field is $\xi=\partial_\phi$ and $\xi^2=r^2$. Quotient space can be identified by $\phi \equiv \phi + 2 n \pi $. The topology of space-time is $\mathbb{R}^3 \cross S^1$. 
 \par
 By introducing the coordinates on hyperplane $\{y_0, y_2, y_3\}$ as
 \begin{eqnarray}
 y_0 &=& f(r) \text{Sinh} \left(\frac{r_+t}{l}\right) \nonumber \\
 y_2 &=& f(r) \text{Cosh} \left(\frac{r_+t}{l}\right) \text{Cos}\left(\frac{r_+\theta}{l}\right) \nonumber \\
 y_2 &=& f(r) \text{Cosh} \left(\frac{r_+t}{l}\right) \text{Sin}\left(\frac{r_+\theta}{l}\right) \ ,
 \end{eqnarray} 
 with $f(r)=l\sqrt{\frac{r^2}{r_+^2}-1}$. Using this the metric can be written as 
 \begin{equation} \label{btz3+1}
     ds^2= -\frac{r^2-r_+^2}{l^2} dt^2 + \frac{l^2}{r^2-r_+^2} dr^2 +(r^2-r_+^2) \text{Cosh}^2\left(\frac{r_+ t}{l^2}\right)d \theta^2+ r^2 d\phi^2 \ .
 \end{equation}
 One can notice that the metric is non-static. The space-time has topology $\mathbb{R}^2 \cross \mathbb{T}^2$. Thus it describes a growing toroidal black hole. By defining $u^2=r^2-r_+^2$ one can easily verify that throat lies at $u=0$. By calculating the  Kretschmann scalar	$K=R_{\alpha \beta \gamma \delta} R^{\alpha \beta \gamma \delta}$, there is a symmetry of both side of the throat. At $u=0$ we have minimum area $4 \pi r_+^2$.
 	      \par
 	      Using embedding in \ref{Kruskal} we can find the metric in Kruskal like $(U,V,\theta, \phi)$ co-ordinate is
 	      \begin{equation} \label{Metric u v}
 	        ds^2 =  \frac{1 }{(1+UV)^2}\Bigg( -4l^2 dUdV + r_+^2 (U-V)^2 d \theta^2 + r_+^2(-1+UV)^2d\phi^2 \Bigg) \ .
 	      \end{equation}
 	       The further quotient of $\mathbb{Z}_2$ will give rise to an isometry $J$ with identification $J : (U, V, \phi, \theta) \rightarrow (V,U , \phi, \theta +\pi)$.  Using the linearized equation at $V=0$, integrating over all $U$ with appropriate AdS boundary conditions reduces to \footnote{Detailed discussion in Appendix \ref{Linearised}.} 
		\begin{equation} \label{huu4}
		 \int dU h_{UU} = \frac{32  \pi G_N l^2}{5}   \int dU T_{UU} \ .
		 \end{equation}
		 To find the shift $\Delta V$ at $U=\infty$ we have
		 \begin{equation}
		     \Delta V (+\infty) = -\frac{32  \pi G_N l^2}{5 g_{UV}(0)} \int_{-\infty}^{\infty} dU \expval{T_{kk}} = \frac{16  \pi G_N }{5 } \int_{-\infty}^{\infty} dU \expval{T_{kk}} \ .
		 \end{equation}
	For the traversability of the wormhole ANEC has to violate i.e., $\Delta V < 0$
	%%%%%%%%%%%%%%%%%%%%%%%%%%%%%%%%%%%%%%%%%%%%%%%%%%%%%%%%%%%%%%%%%%%%%%%%%%%%%%%%%%%%%%%%%%%%%%%%%%%%%%%%%%%%%%%%%%
	\section{The scalar field} \label{scalar}
	  
	  From the above discussion, to examine the ANEC, we have to compute the expectation value of stress energy tensor. As the expectation value is evaluated in covering space $\tilde{M}$ but due to the property that it is the quotient of AdS$_4$ with identification $\phi \thicksim \phi+2n \pi$. From equation \eqref{expac} by using two point function one can compute the expectation value of stress-energy tensor.
		The scalar two-point function in arbitrary dimension $d$ has been discussed in \cite{Allen}. We will quickly summarise the aspects of their results that are relevant to our goal in this section.
		\par 
		The scalar two point function can be written as 
\begin{equation}
    G(x, x') = \bra{\psi}\phi(x)\phi(x')\ket{\psi} \ .
\end{equation}
Here the state $\psi$ is maximally symmetric, $G(x, x')$ solely depends on the geodesic distance $\mu(x, x')$ for spacelike separated points $x, x'$. Since $\mu(x, x')$ is the proper distance along a geodesic, the vectors $n_\alpha(x,x')=\nabla _{\alpha} \mu (x,x')$ and $n_{\alpha '}(x,x')=\nabla _{\alpha '} \mu (x,x')$ have unit length. Because they are pointing away from one another by the relation $n_\alpha  \tensor{g}{^\alpha_{\beta '}} = - n_{\beta '}$. The parallel propagator $g_{\alpha \beta '}(x, x' )$ along the geodesic joining $x$ to  $x'$ is unique for maximally symmetric spaces. It possesses the following properties:
\begin{eqnarray}
\tensor{g}{_{\alpha \beta '}} (x, x') &=& \tensor{g}{_{\alpha \beta }} (x) \hspace{0.2cm}\text{for} \hspace{0.2cm} x=x' \\
\tensor{g}{_{\alpha \beta '}} (x, x') &=& \tensor{g}{_{ \beta ' \alpha  }} (x', x) \\
g_{\alpha \beta}(x) &=& g_{\alpha \gamma '}(x, x') g_{\beta \delta '} (x, x') g^{\gamma ' \delta '}(x') \ .
\end{eqnarray}
The derivatives of $n_\alpha$ and $g_{\alpha \beta}$ may thus be represented in terms of our fundamental set:
\begin{eqnarray} \label{danb}
\nabla _\alpha n_\beta &=& A(\mu) \left[ g_{\alpha \beta} (x) - n_\alpha (x, x') n_\beta (x, x') \right] \\ \label{danb'}
\nabla _\alpha n_{\beta '} &=& C(\mu) \left[ g_{\alpha {\beta '}} (x, x') +  n_\alpha (x, x') n_{\beta '}(x, x') \right] \\
\nabla _\alpha g_{\beta \gamma '} &=& -\left[A(\mu) + C(\mu)\right] \{ g_{\alpha \beta}(x) n_{\gamma '}(x, x') +  g_{\alpha \gamma '} (x, x') n_\beta (x, x') \} \ .
\end{eqnarray}
and the similar expression for the derivatives with respect to $\nabla_{\alpha '}$. The functions $A(\mu)$ and $C(\mu)$ are 
\begin{equation}
    A(\mu) = \frac{1}{l} \text{Coth} \left(\frac{\mu}{l}\right) \hspace{2cm} C(\mu) = -\frac{1}{l} \text{Cosech} \left(\frac{\mu}{l}\right) \ .
    \end{equation}
     \par
     
     By using $G' = \frac{dG}{d\mu}$, we obtain 
     \begin{eqnarray}
     \Box G(\mu) &=& \nabla^\alpha \nabla_\alpha G (\mu) \nonumber \\
     &=& \nabla^\alpha(G'(\mu) n_\alpha) \hspace{0.5cm}  \nonumber \\
     &=& G''(\mu)  + 3 G'(\mu) A(\mu) \ .
     \end{eqnarray}
     In the second line we have used $ \nabla_\alpha \mu = n_\alpha $ and in last $\tensor{\delta}{_\alpha^\alpha}=4$ and $n^\alpha n_\alpha =1$.
     The equation of motion $(\Box-m^2)\phi(x)=0$ can be written as (if $x \neq x')$,
     \begin{equation} \label{EOM}
         G''(\mu)  + 3 G'(\mu) A(\mu) -m^2 G=0 \ .
     \end{equation}
     By changing of variable as 
     \begin{equation}
         z= \text{Cosh}^2 \left(\frac{\mu}{2l}\right) \ ,
     \end{equation}
    then the equation \eqref{EOM} reduces to  
	\begin{equation} \label{hypergeo}
	\left[z(1-z) \frac{d^2}{dz^2}+\{c-(a_++a_-+1)\} \frac{d}{dz} - a_+a_-\right] G(z) = 0 \ ,
	\end{equation}
	with  the parameters $ 2a_\pm = \left[3\pm \sqrt{9 + 4 m^2 l^2}\right]$ and $c=2$.
	The two point function G(z) as the solution of \eqref{hypergeo} in terms of hypergeometric functions can be written as 
	\begin{equation} \label{Gz}
	G(z) = p z^{-a_+} F(a_+, a_+-c+1; a_+-a_-+1; z^{-1}) \ ,
	\end{equation}
	with normalization constant $p$ is given by
	\begin{equation} \label{r}
	p =  \frac{\Gamma(a_+) \Gamma(a_+-c+1)}{16 \pi^2 l^2 \Gamma(a_+-a_-+1)}  \ .
	\end{equation} 
	In the case of AdS$3$ the hypergeometric function summed up to a exact analytic expression. But in the case of AdS$_4$, we don't have the exact analytic form. In the terms of the conformal weight $\Delta=\sqrt{9+4 m^2 l^2}$, the above parameters reduces to $2a_\pm = 3 \pm \Delta$ and $c=2$.
		\par
			A standard calculation using the embedding in equation \eqref{Embedding} the geodesics distance is
			\begin{eqnarray} \label{geodesics}
	\mu &=& \frac{l^2}{(1+UV)(1+U'V')} \Bigg[ (U+V)(U'+V')- (U-V)(U'-V') \text{Cos} \left(\frac{r_+}{l}(\theta-\theta')\right) \nonumber \\
	&& + (-1+U'V')(-1+UV)\text{Cosh}\left(\frac{r_+}{l}(\phi-\phi')\right) \Bigg]  \nonumber \ . 
	\end{eqnarray}
	By defining $\mathcal{C}=\text{Cos} \left(\frac{r_+}{l}(\theta-\theta')\right)$ and $\mathcal{K}=\text{Cosh}\left(\frac{r_+}{l}(\phi-\phi')\right)$ the above expression can be written as 
	\begin{eqnarray}
	\mu = \frac{l^2}{(1+UV)(1+U'V')} \Bigg[ (U+V)(U'+V')- (U-V)(U'-V') \mathcal{C} + (-1+U'V')(-1+UV)\mathcal{K} \Bigg] \nonumber \ .
	\end{eqnarray}
			Working on the horizon $V=0$, we define
		\begin{equation} \label{f}
		f(\mathcal{K},U,\mathcal{C},\Delta) = \langle0 |\partial_U \phi(x) \partial_U \phi(x') |0\rangle \ ,
		\end{equation}
		with $x=(U, V, \theta, \phi)$ and $x'=(U', V', \theta ', \phi ')$ points in AdS$_4$

		%%%%%%%%%%%%%%%%%%%%%%%%%%%%%%%%%%%%%%%%%%%%%%%%%%%%%%%%%%%%%%%%%%
%%%%%%%%%%%%%%%%%%%%%%%%%%%%%%%%%%%%%%%%%%%%%%%%%%%%%%%%%%%%%%%%%%%
         \subsection{Calculations of Stress Energy tensor}
       As we claimed above, we don't have closed form for two point function in the case of AdS$_4$. We will compute two point function for different values of $m^2$. From equation \eqref{Gz} for different values of $m^2$ we will have two point function. For $(m^2=0)$
    \begin{eqnarray} \label{Gz for zero}
        G(z) &=& \frac{\Gamma(3) \Gamma(2)}{\Gamma(4) 16 \pi^2} z^{-3} F\left(3, 2, 4, \frac{1}{z} \right) \nonumber \\
            &=& \frac{1}{16 \pi^2 z(z-1)}\left[-1+2z+(2z^2-2z)\text{log}\left(\frac{z-1}{z}\right)\right] \ .
    \end{eqnarray}
		
	\begin{figure}[ht]
\begin{subfigure}{.5\textwidth}
  \centering
  % include first image
  \includegraphics[width=.8\linewidth]{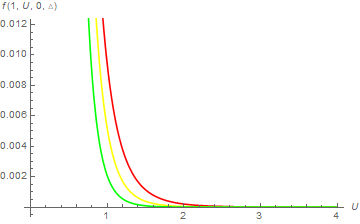}  
  %\caption{Put your sub-caption here}
  \label{fig:sub-first}
\end{subfigure}
\begin{subfigure}{.5\textwidth}
  \centering
  % include second image
  \includegraphics[width=.8\linewidth]{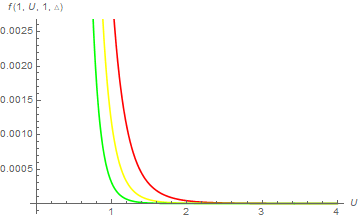}  
  %\caption{Put your sub-caption here}
  \label{fig:sub-second}
\end{subfigure}
\caption{Some of the functions \textbf{Left:}  $\mathcal{K}=1,  \mathcal{C}=0$,  for $\Delta=3$ (red), $\Delta=5$ (yellow), $\Delta=7$ (green). \textbf{Right:} $\mathcal{K}=1, \mathcal{C}=0$,  for $\Delta=3$ (red), $\Delta=5$ (yellow), $\Delta=7$ (green).}
\label{Ads4 1}
\end{figure}
		
		\begin{figure}[ht]
\begin{subfigure}{.5\textwidth}
  \centering
  % include first image
  \includegraphics[width=.8\linewidth]{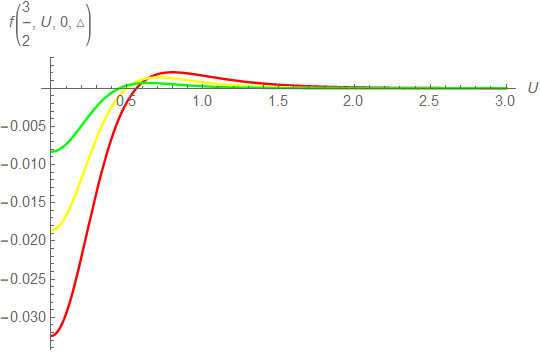}  
  %\caption{Put your sub-caption here}
  \label{fig:sub-first}
\end{subfigure}
\begin{subfigure}{.5\textwidth}
  \centering
  % include second image
  \includegraphics[width=.8\linewidth]{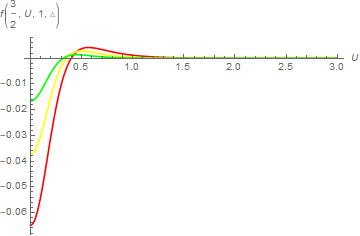}  
  %\caption{Put your sub-caption here}
  \label{fig:sub-second}
\end{subfigure}
\caption{Some of the functions \textbf{Left:}  $\mathcal{K}=1.5, \mathcal{C}=0$, for $\Delta=3$ (red), $\Delta=5$ (yellow), $\Delta=7$ (green). \textbf{Right:}  $\mathcal{K}=1.5, \mathcal{C}=0$,  for $\Delta=3$ (red), $\Delta=5$ (yellow), $\Delta=7$ (green).}
\label{Ads4 2}
\end{figure}	
	The graphs are identical as \cite{Scalar} for BTZ in $2+1$ case. Now computation of $f$ for $m^2=0$ we have
\begin{equation*}
    f(\mathcal{C},U,\mathcal{K}) = \frac{(1+\mathcal{C})\left[\mathcal{K}-\mathcal{K}^2l^2+(1+\mathcal{C})(-1+2\mathcal{K}l^2)U^2+3(1+\mathcal{C})^2l^2U^4\right]}{16 l^2 \pi^2\left(\mathcal{K}+(1+\mathcal{C})U^2\right)^3 \left(-1+l^2(\mathcal{K}+(1+\mathcal{C})U^2\right)^3} \ .
\end{equation*}
By fixing $\mathcal{C}=1$ and $l=1$ the above expression reduces to
\begin{equation*}
   	f(U,\mathcal{K}) = \frac{\mathcal{K}-\mathcal{K}^2+(-1+2\mathcal{K})U^2+3U^4}{16 \pi^2 (-1+\mathcal{K}+U^2)^3(\mathcal{K}+U^2)^3} \ . \ .
\end{equation*}
By integrating the above equation i.e.,
\begin{equation}
    \int_0^\infty f(U,\mathcal{K}) dU = - \frac{\sqrt{1-\mathcal{K}}+3\mathcal{K}\sqrt{1-\mathcal{K}}-4\mathcal{K}^2\sqrt{1-\mathcal{K}}}{128 \pi (\mathcal{K}(1-\mathcal{K})){3/2}} < 0  \ .
\end{equation}
This shows the violation of ANEC. By plotting $\int_{0}^{\infty} f(\mathcal{K}, U, \mathcal{C}, \Delta) $ for fixed value of $\mathcal{C}$ (e.g $0,1$) with respect to $\mathcal{K}$ one can also verify the same.
	
			\begin{figure}[ht]
\begin{subfigure}{.5\textwidth}
  \centering
  % include first image
  \includegraphics[width=.8\linewidth]{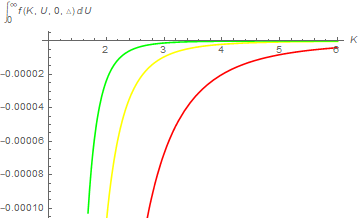}  
  %\caption{Put your sub-caption here}
  \label{fig:sub-first}
\end{subfigure}
\begin{subfigure}{.5\textwidth}
  \centering
  % include second image
  \includegraphics[width=.8\linewidth]{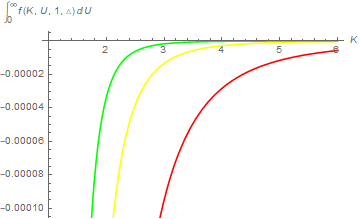}  
  %\caption{Put your sub-caption here}
  \label{fig:sub-second}
\end{subfigure}
\caption{\textbf{Left:} $\int_{0}^{\infty} f(\mathcal{K}, U, \mathcal{C}, \Delta) $  for $\Delta=3$ (red), $\Delta=5$ (yellow), $\Delta=7$ (green). \textbf{Right:}  $\int_{0}^{\infty} f(\mathcal{K}, U, \mathcal{C}, \Delta) $  for $\Delta=3$ (red), $\Delta=5$ (yellow), $\Delta=7$ (green).}
\label{Ads4 3}
\end{figure}

		 %%%%%%%%%%%%%%%%%%%%%%%%%%%%%%%%%%%%%%%%%%%%%%%%%%%%%%%%%%%%%%%%
  \section{Results and Discussion}
%%%%%%%%%%%%%%%%%%%%%%%%%%%%%%%%%%%%%%%%%%%%%%%%%%%%%%%%%%%%%%%%%%%%
In this paper, we posed the issue of wormhole traversability in a quotient of the space-time obtained from quotients of $AdS_4$ space with specific $\mathbb{Z}_2$ symmetry in the presence of scalar fields. Back-reaction from quantum scalar fields in Hartle-Hawking states is explored on simple explicit examples of $\mathbb{Z}_2$ wormholes asymptotic to $ \rm AdS_4$. These examples are often traversable when the scalar satisfies periodic boundary condition (choosen from eq\eqref{PAP condition}) around the $\mathbb{Z}_2$ cycle. In $AdS_4$, we found the expression for the scalar fields's two-point function. We calculated the average null energy using this and discovered that it becomes negative when the periodic boundary conditions on the scalar fields are chosen. The wormholes can then be traversable due to the back reaction on the geometry. It's also interesting extrapolating this to investigate wormhole traversability in the context of higher spin fields. Recent research has revealed that Euclidean wormholes serve an important role in providing a new viewpoint on the information loss paradox. It would be fascinating to see if the problem of traversability in Lorentzian wormholes like the ones investigated in this paper sheds any light on this topic. For static space-time, Hartle-Hawking Vacua has been discussed in \cite{Jacobson:1994fp}. We're using Hartle-Hawking vacua to calculate expectation values, but their forms aren't essential. It would be fascinating to learn the form of the Hartle-Hawking state for non-static space-times, particularly for the above metric.
		%%%%%%%%%%%%%%%%%%%%%%%%%%%%%%%%%%%%%%%%%%%%%%%%%%%%%%%%%%%%%%%%%%%%%%%%%%%%%%%%%%%%%%%%%%%%%%%%%%%%%%%%%%%%%%%%%%%%%%%%

\section*{Acknowledgement}

I am indebted to Prasanta K. Tripathy for many helpful discussions as well as for a careful manuscript reading.

%%%%%%%%%%%%%%%%%%%%%%%%%%%%%%%%%%%%%%%%%%%%%%%%%%%%%%%%%%%%%%%%%%%%%%%%%%%%%%%%%%%%%%%%%%%%%%%%%%%%%%%%%%%%%%%%%%%%%%%%%%%%%%%%%%%%%%%
		\begin{appendices}

\section{Co-ordinate Choice and their Kruskal Extension}\label{Kruskal}
The choice of co-ordinate is
 \begin{eqnarray} \label{coordinate choice}
 T_1 &=& \frac{r l}{r_+} \text{Cosh} \left(\frac{r_+ \phi}{l} \right) \nonumber \\
 T_2 &=& f(r) \text{Sinh} \left(\frac{r_+ t}{l^2} \right) \nonumber \\
 X_1 &=& \frac{r l}{r_+} \text{Sinh} \left(\frac{r_+ \phi}{l} \right) \nonumber \\
 X_2 &=& f(r) \text{Cosh} \left(\frac{r_+ t}{l^2} \right) \text{Cos} \left(\frac{r_+ \theta}{l}\right) \nonumber \\
  X_3 &=& f(r) \text{Cosh} \left(\frac{r_+ t}{l^2} \right) \text{Sin} \left(\frac{r_+ \theta}{l}\right)   \ . 
 \end{eqnarray}
 Again by using equation \eqref{Hyperboloid} one can find $f(r)=l \sqrt{\frac{r^2}{r_+^2}-1}$.
For the Kruskal extension, one can start with
\begin{eqnarray}
ds^2 &=& -\frac{r^2-r_+^2}{l^2} dt^2 + \frac{l^2}{r^2-r_+^2} dr^2 \nonumber \\ \label{rtonly}
    &=& \frac{r^2-r_+^2}{l^2} \left[ -dt^2+dr_*^2\right] \ .
\end{eqnarray}
One can find $r_* = \frac{l^2}{2r_+}\text{ln} \frac{|r-r_+|}{r+r_+}$. Lets define the co-ordinate $u=t-r_*$ and $v=t+r_*$ then the metric \eqref{rtonly} can be written as 
\begin{equation}
    ds^2 = \frac{r^2-r_+^2}{l^2} \left[ -du dv \right] \ .
\end{equation}
As we have $t=\frac{u+v}{2}$ and $r_*=\frac{v-u}{2}$ then 
\begin{eqnarray}
\frac{r^2-r_+^2}{l^2} &=& \frac{(r+r_+)^2}{l^2} e^{\frac{r_+}{l^2}(v-u)} \ .
\end{eqnarray}
Let's define again $U=-e^{-\frac{r_+}{l^2}u}$ and $V=e^{\frac{r_+}{l^2}v}$, using this $t=- \frac{l^2}{2r_+} \text{ln} \left( - \frac{U}{V}\right)$ and $r=r_+ \frac{1-UV}{1+UV}$, and finally our co-ordinate becomes 
\begin{eqnarray} \label{Embedding}
T_1 &=& l \frac{1-UV}{1+UV} \text{Cosh} \left( \frac{r_+}{l} \phi \right) \nonumber \\
T_2 &=& l \frac{V+U}{1+UV} \nonumber\\
X_1 &=& l \frac{1-UV}{1+UV} \text{Sinh} \left( \frac{r_+}{l} \phi \right) \nonumber \\
X_2 &=& l \frac{V-U}{1+UV} \text{Cos} \left( \frac{r_+}{l} \theta \right) \nonumber\\
X_3 &=& l \frac{V-U}{1+UV} \text{Sin} \left( \frac{r_+}{l} \theta \right) \ .
\end{eqnarray}
By using the formula 
\begin{equation}
   \mu(U,V,\phi,\theta,U',V',\phi',\theta')= T_1 T_1' + T_2 T_2' - X_1 X_1' - X_2 X_2' - X_3 X_3' \ ,
\end{equation}
one can get the form of geodesics distance same as \eqref{geodesics}.
%%%%%%%%%%%%%%%%%%%%%%%%%%%%%%%%%%%%%%%%%%%%%%%%%%%%%%%%%%%%%%%%%%%%%%%%%%%%%%%%%%%%%%%%%%%%%%%%%%%%%%%
%%%%%%%%%%%%%%%%%%%%%%%%%%%%%%%%%%%%%%%%%%%%%%%%%%%%%%%%%%%%%%%%%%%%%%%%%%%%%%%%%%%%%%%%%%%%%%%%%%%%%%%%%%
\section{Linearized Equation} \label{Linearised}
In this section we derive the Linearized Einstein in both case e.g., AdS$_3$ and AdS$_4$ case. 
     \subsection{AdS$_3$ case}
     The Einstein equation in the $\rm AdS_3$ case can be written as 
     \begin{equation}
     R_{\alpha \beta} - \frac{1}{2}g_{\alpha \beta}\left(R+\frac{2}{l^2} \right) = 8 \pi G_N T_{\alpha \beta} \ .
     \end{equation}
     With the small perturbation in the metric i.e., $g_{\alpha \beta}=g_{\alpha \beta}+\delta g_{\alpha \beta}=g_{\alpha \beta}+\epsilon h_{\alpha \beta} + \order{\epsilon ^2}$. We are perturbing the metric in Kruskal like co-ordinate. By putting this in Einstein equation and putting $V=0$ as we are working on $V=0$ horizon. Only taking the terms independent of $\frac{1}{U}$ as we are interested in $U \rightarrow \pm \infty$
     \begin{eqnarray}
         \frac{1}{2l^2} \left[2 h_{UU}+U \partial _U h_{UU} - \frac{1}{2r_+^2} \partial_U ^2 h_{\phi \phi}\right] &=& 8 \pi G_N T_{UU} \nonumber \\
         \frac{1}{2l^2} \left[2 h_{UU}+ \partial _U (U h_{UU}) - h_{UU} - \frac{1}{2r_+^2} \partial_U ^2 h_{\phi \phi}\right] &=& 8 \pi G_N T_{UU} \nonumber \\
        \frac{1}{2l^2} \left[ h_{UU}+ \partial _U (U h_{UU}) - \frac{1}{2r^2} \partial_U ^2 h_{\phi \phi}\right] &=& 8 \pi G_N T_{UU} \ .
     \end{eqnarray}
     By integrating over all $U$ and dropping the boundary terms as the requirements of boundary stress tensor be unchanged at this order. Finally we have 
     \begin{equation}
         \int dU h_{UU} = 16\pi G_N l^2 \int dU T_{UU}
     \end{equation}
     %%%%%%%%%%%%%%%%%%%%%%%%%%%%%%%%%%%%%%%%%%%%%%%%%%%%%%%%%%%%%%%%%%%%%%%%%%%%%%%%%%%%%%%%%%%%%%%%%%%
     \subsection{AdS$_4$ case}
      The Einstein equation in the $\rm AdS_4$
     \begin{equation}
     R_{\alpha \beta} - \frac{1}{2}g_{\alpha \beta}\left(R+\frac{6}{l^2} \right) = 8 \pi G_N T_{\alpha \beta} \ .
     \end{equation}
     Again perturbing the metric and taking the limits as above it is easy to verify that we have 
     \begin{eqnarray}
         \frac{1}{2l^2}\left[  4 h_{UU}+  \frac{3}{2} U\partial _U h_{UU} - \frac{1}{2r_+^2} \partial_U ^2 h_{\phi \phi}\right] &=& 8 \pi G_N T_{UU} \nonumber \\
         \frac{1}{2l^2}\left[  4 h_{UU}+  \frac{3}{2} \left \{ \partial _U (U h_{UU})-h_{UU} \right \} - \frac{1}{2r_+^2}  \partial_U ^2 h_{\phi \phi}\right] &=& 8 \pi G_N T_{UU} \nonumber \\
        \frac{1}{2l^2} \left[\frac{5}{2} h_{UU}+  \frac{3}{2}\partial _U (U h_{UU}) - \frac{1}{2r^2} \partial_U ^2 h_{\phi \phi}\right] &=& 8 \pi G_N T_{UU} \ .
     \end{eqnarray}
     By integrating over all $U$ and dropping the boundary terms as the requirements of boundary stress tensor be unchanged at this order. Finally we have 
     \begin{equation}
         \int dU h_{UU} = \frac{32 \pi G_N l^2}{5} \int dU T_{UU}
     \end{equation}
     This is same as eq\eqref{huu4}
%%%%%%%%%%%%%%%%%%%%%%%%%%%%%%%%%%%%%%%%%%%%%%%%%%%%%%%%%%%%%%%%%%%%%%%%%%%%%%%%%%%%%%%%%%%%%%%%%%%%%%%%%%%%%%%%%%%%%%%%%%%%%%5%%%%

\end{appendices}

%%%%%%%%%%%%%%%%%%%%%%%%%%%%%%%%%%%%%%%%%%%%%%%%%%%%%%%%%%%%%%%%%%%%%%%%%%%%%%%%%%%%%%%%%%%%%%%%%%%%%%%%%%%%%%%%%%%%%%%%%%%%%%%%%%%%%%%%

%%%%%%%%%%%%%%%%%%%%%%%%%%%%%%%%%%%%%%%%%%%%%%%%%%%%%%%%%%%%%%%%%%%%%%%%%%%%%%%%%%%%%%%%%%%%%%%%%%%%%%%%%%%%%%%%%%%%%%%%%%%%%%%%%%%%%%%%

\end{document}